\documentclass[sigconf]{acmart}

\usepackage{enumitem}
\usepackage{xspace}
\usepackage{multirow}
\usepackage{bm}
\usepackage{subfigure}

\newcommand{\sys}{\textsc{PROMO}\xspace}

\AtBeginDocument{%
  }

\setcopyright{acmlicensed}
\copyrightyear{2018}
\acmYear{2018}
\acmDOI{XXXXXXX.XXXXXXX}

\acmConference[Conference acronym 'XX]{Make sure to enter the correct
  conference title from your rights confirmation emai}{June 03--05,
  2018}{Woodstock, NY}
\acmISBN{978-1-4503-XXXX-X/18/06}

\copyrightyear{2024}
\acmYear{2024}
\setcopyright{acmlicensed}\acmConference[RecSys '24]{18th ACM Conference on Recommender Systems}{October 14--18, 2024}{Bari, Italy}
\acmBooktitle{18th ACM Conference on Recommender Systems (RecSys '24), October 14--18, 2024, Bari, Italy}
\acmDOI{10.1145/3640457.3688126}
\acmISBN{979-8-4007-0505-2/24/10}

\begin{document}

\title{Prompt Tuning for Item Cold-start Recommendation}

\author{Yuezihan Jiang}
\affiliation{
  \institution{Kuaishou Technology}
  \city{Beijing}
  \country{China}
  }
\email{yuezihan.jiang@gmail.com}

\author{Gaode Chen}
\authornote{Corresponding author.}
\affiliation{
  \institution{Kuaishou Technology}
  \city{Beijing}
  \country{China}
  }
\email{chengaode19@gmail.com}

\author{Wenhan Zhang}
\affiliation{
  \institution{Peking University}
  \city{Beijing}
  \country{China}
}
\email{pku.zwh@pku.edu.cn}

\author{Jingchi Wang}
\affiliation{
  \institution{Peking University}
  \city{Beijing}
  \country{China}
}
\email{wangjingchi@pku.edu.cn}

\author{Yinjie Jiang}
\affiliation{%
  \institution{Kuaishou Technology}
  \city{Beijing}
  \country{China}
}
\email{jiangyinjie@kuaishou.com}

\author{Qi Zhang}
\affiliation{%
  \institution{Kuaishou Technology}
  \city{Beijing}
  \country{China}
}
\email{zhangqi38@kuaishou.com}

\author{Jingjian Lin}
\affiliation{%
  \institution{Kuaishou Technology}
  \city{Beijing}
  \country{China}
}
\email{linjingjian@kuaishou.com}

\author{Peng Jiang}
\affiliation{%
  \institution{Kuaishou Technology}
  \city{Beijing}
  \country{China}
}
\email{jp2006@139.com}

\author{Kaigui Bian}
\authornotemark[1]
\affiliation{%
  \institution{Peking University}
  \city{Beijing}
  \country{China}
}
\email{bkg@pku.edu.cn}

\renewcommand{\shortauthors}{Jiang et al.}

\begin{abstract}
The item cold-start problem is crucial for online recommender systems, as the success of the cold-start phase determines whether items can transition into popular ones. 
Prompt learning, a powerful technique used in natural language processing (NLP) to address zero- or few-shot problems, has been adapted for recommender systems to tackle similar challenges. 
However, existing methods typically rely on content-based properties or text descriptions for prompting, which we argue may be suboptimal for cold-start recommendations due to 1) semantic gaps with recommender tasks, 2) model bias caused by warm-up items contribute most of the positive feedback to the model, which is the core of the cold-start problem that hinders the recommender quality on cold-start items. 
We propose to leverage high-value positive feedback, termed pinnacle feedback as prompt information, to simultaneously resolve the above two problems. 
We experimentally prove that compared to the content description proposed in existing works, the positive feedback is more suitable to serve as prompt information by bridging the semantic gaps. 
Besides, we propose item-wise personalized prompt networks to encode pinnaclce feedback to relieve the model bias by the positive feedback dominance problem. 
Extensive experiments on four real-world datasets demonstrate the superiority of our model over state-of-the-art methods. Moreover, \sys has been successfully deployed on a popular short-video sharing platform, a billion-user scale commercial short-video application, achieving remarkable performance gains across various commercial metrics within cold-start scenarios.
\end{abstract}





\maketitle

\section{Introduction}
Recently, item cold-start recommendation has garnered significant attention from researchers~\cite{liu2023uncertainty,cao2023multi}, as every item inevitably undergoes this phase, which significantly shapes its potential for popularity.
With the burgeoning success of pre-training techniques in Natural Language Processing (NLP) \cite{min2023recent}, numerous endeavors have been made to integrate pre-training into cold-start recommendations \cite{hao2023multi,hao2021pre}.

These methods leverage pre-trained knowledge to alleviate the sparsity issue, thereby enhancing the quality of cold-start recommendations \cite{xiao2021uprec}. 
However, amidst the prevailing trend of pre-training, the effective extraction of useful information from pre-trained models emerges as a promising avenue. 
One such method, prompt-tuning \cite{liu2022p,lester2021power}, stands out for its remarkable advantages over classical fine-tuning paradigms, particularly in zero-shot and few-shot scenarios. 
By employing hard text templates or soft continuous embeddings as prompts, prompt-tuning transforms downstream tasks into analogous well-trained pre-training tasks.

The merits of prompt-tuning lie in two key aspects: firstly, it bridges the gap between pretraining and downstream objectives, thereby optimizing the utilization of pretraining model knowledge, which is especially advantageous in cold-start scenarios. 
Secondly, prompt-tuning requires tuning only a small set of parameters for the prompts and labels, making it more parameter-efficient.

Although researchers have applied prompt learning to address cold-start recommendations by reframing them as zero- or few-shot problems \cite{wu2024personalized,yi2023contrastive}, the effectiveness of prompt learning in real industrial recommendation settings remains suboptimal from both data and model perspectives.

\begin{itemize}[leftmargin=*]
    \item \textbf{Data side cost and gap}. Existing approaches often necessitate additional human annotations, such as text prompt descriptions for candidate items, resulting in significant practical costs \cite{zhai2023knowledge}.
    Moreover, relying solely on content features as prompt information may not be well-suited for recommender tasks, lacking an end-to-end connection to the recommendation process.
    \item \textbf{Model side bias}.
     A critical challenge in item cold-start recommendations arises from the poor model performance on cold-start items due to limited positive feedback contribution. 
     This disparity often leads to situations where the model assigns high scores to warm-start items and low scores to cold-start items.
    \cite{zhai2023knowledge,geng2022recommendation} employ a shared model architecture for prompt encoding, potentially resulting in insufficient personalization for cold-start items. 
    This is because cold-start items represent only a small fraction of online traffic compared to popular items, causing the model parameters to be primarily optimized for popular items. 
    Besides, the length of the prompt embedding restricts the volume of tunable parameters, and the prompt information is concentrated in the input layer, leading to less impact on the model.
\end{itemize}

\begin{figure}
	\centering
	\includegraphics[width=0.8\columnwidth]{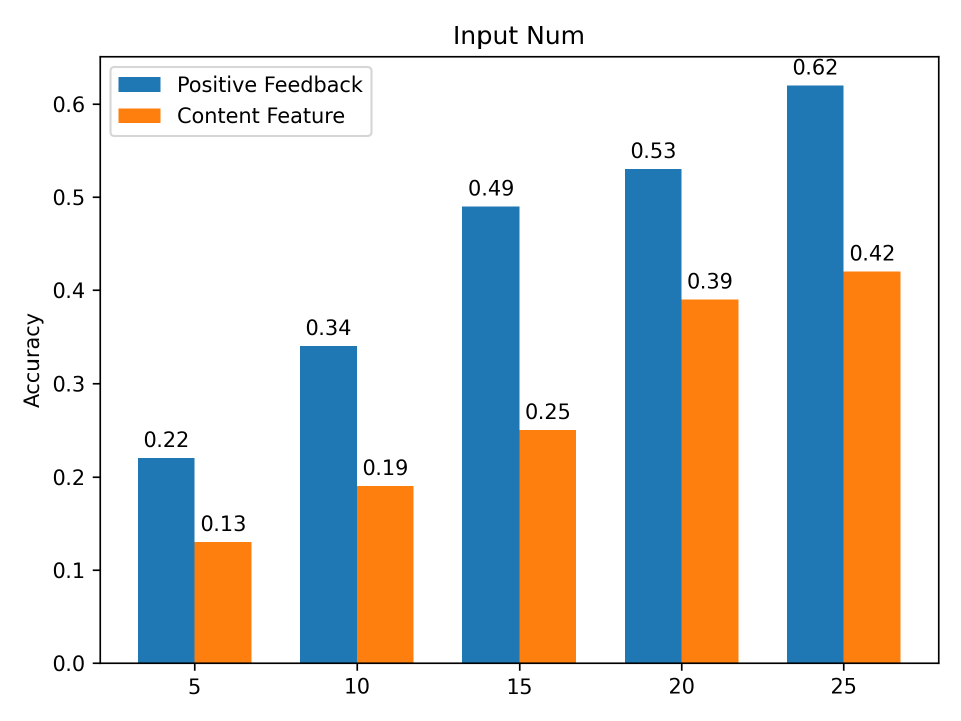}
	\caption{The model accuracy on cold-start items with item-side taking different inputs for representations encoding. The positive feedback provides more task-relevant information as the model with positive feedback information gains higher accuracy on user preference.}
	\label{case1}

\end{figure}
\begin{figure}
	\centering
	\includegraphics[width=0.8\columnwidth]{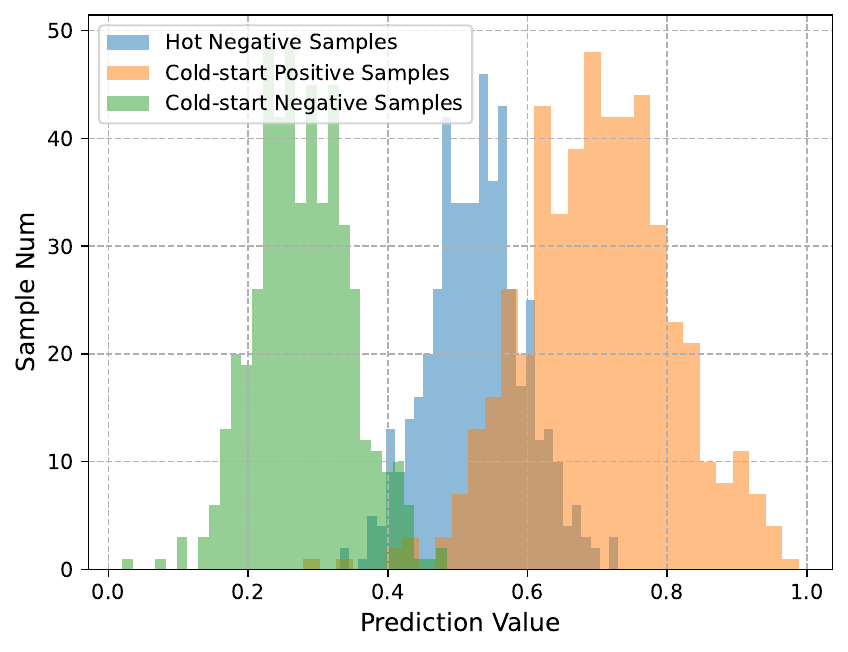}
	\caption{The histogram of prediction values on cold-start items and warm-start items. The X-axis represents the prediction value, and the Y-axis represents the sample amount of the corresponding range of scores.}
	\label{case2}

\end{figure}

To address the data side problem, we propose leveraging positive feedback as effective prompt data for cold-start items.
This choice is motivated by the direct relevance of positive feedback to recommender tasks.
To verify the relevance of positive feedback to downstream recommender tasks, we pretrain two dual-tower modelsl~\cite{huang2013learning} on KuaiRand-Pure~\cite{gao2022kuairand}, utilizing all user features.
The only distinction between the models lies in the input from the item side: one model takes item features as input, while the other takes the user IDs of historical positive feedback users. 
We sample 300 cold-start items (viewed less than 1,000 times by users in KuaiRand) to examine the performance of the two models. 
The X-axis in Figure~\ref{case1} represents the number of content features or IDs from the positive feedback users, and the Y-axis corresponds to the model's accuracy in predicting whether a user will click on an item. 
It's evident that the model trained with positive feedback demonstrates superior performance across all settings. 
This is attributed to ID embeddings being infused with user interest information during training. 
As the IDs space is continuous, they provide more informative signals than discrete content features, aiding the model in making more accurate predictions regarding user preferences.

From a model perspective, a critical challenge in item cold-start recommendations arises from poor performance on cold-start items due to limited positive feedback contribution. 
This leads to a situation where the model assigns high scores to warm-start items and low scores to cold-start items. 
In Figure~\ref{case2}, histograms of prediction scores on 500 positive and 500 negative feedback on cold-start items, along with 500 negative feedback on warm-start items, are plotted separately from the online sorting model in the pretrained dual-tower model on KuaiRand dataset. 
It's observed that the mean values of the prediction scores on cold-start positive samples and warm-start negative samples are very close to each other. 
Additionally, the overlap of the prediction distributions of cold-start positive samples and warm-start negative samples is much larger than the overlap of the predictions on cold-start positive samples and cold-start negative samples, indicating that the model parameters are predominantly influenced by warm-start items, resulting in limited personalization for cold-start items and a tendency to assign higher scores to popular items.

To address these challenges, we present a prompt-tuning method specifically designed for item cold-start recommendation, namely \sys. 
To avoid data side cost and gap, we propose using significant positive feedback---which we term ``pinnacle feedback''---as the prompt information.
To mitigate the model side bias, we suggest constructing personalized prompt networks for each item, thereby preventing model bias towards hot items. 
Additionally, to enhance the pinnacle feedback based prompt tuning and to ensure fair model predictions on items, we introduce positive feedback prompt-enhanced loss and fairness-aware prompt-enhanced loss separately. 
These enhancements aim to improve the performance of cold-start recommendations.

The contributions of this paper are summarized as follows: 
\begin{itemize}[leftmargin=*]
\item {\it New challenges:} We identify challenges in deploying an industrially efficient prompt-tuning recommendation, including the data cost and gap (from the data perspective) and the limited personalization for cold-start items (from the model perspective).
\item {\it New technical solutions:} \sys makes the initial attempt to address both the challenges: (1) to avoid extra annotation cost and boost recommender performance, \sys suggests utilizing pinnacle feedback for cold-start items as prompts and introduces soft prompts for flexible application in downstream tasks; (2) To avoid model bias caused by positive feedback dominance by warm-start items and improve the recommendation quality on cold-start items, we propose a personalized prompt network to adaptively update item representations. 
Furthermore, we introduce two types of prompt-enhanced loss to boost the performance of cold-start recommendations.
\item {\it SOTA performance:} Extensive experiments on four benchmark recommender datasets and deployment of \sys in a real-world platform at Kuaishou demonstrate that \sys consistently achieves state-of-the-art performance compared to existing baseline methods in terms of a series of recommender statistics.
\end{itemize}

\section{Related Works}
\subsection{Cold-start Recommendation}
In this section, we briefly introduce several hybrid recommender systems that can handle the item cold-start scenario.
Item cold-start recommendation aims to provide effective recommender results to users on the cold-start items. 
It can be divided into several kinds. 

Content-based methods propose to model the distribution of content features, e.g., taking audio information to encode recommender embeddings~\cite{van2013deep,bogdanov2013semantic,chen2021learning}, or taking domain-related information for downstream tasks~\cite{alam2018domain,alam2021corporate,deldjoo2018content,chen2023win}. 
Though content information relieves the data sparsity problem, it is hard to ensure the content information can always supplement the downstream tasks, because there exists a gap between the semantic information and the recommender tasks. 
Collaborative filtering strategies such as CBF and CF have limited scope in the cold start scenario due to lack of information, e.g., involving social relations for collaborative filtering 
~\cite{sedhain2014social,ijcai2021p197}, or proposing a new similarity measure for neighborhood based CF for item cold-start scenarios~\cite{wei2017collaborative}.
Meta-learning addresses the cold-start problem by learning to quickly adapt a model to new items or users with minimal data.
For example,~\cite{zhu2021learning} introduces a meta-learning strategy for item cold-start recommendations by employing deep neural networks that adapt a neural network's biases using item history.
~\cite{lu2020meta} enhances tasks with multifaceted semantic contexts and employs a co-adaptation meta-learner to effectively address the challenges of new item recommendation.
Transfer learning-based recommendation optimizes the cold-start recommendation quality by utilizing cross-domain knowledge. 
For example,~\cite{zhang2021model} learns to transfer model knowledge from rich data settings to few data settings by a meta-learner learning the model parameter shifting relations. 
~\cite{wu2024personalized} utilize prompt learning mode to utilize pretrained knowledge to boost cold-start recommendations. 
\sys belongs to this line of methods. 

\subsection{Prompt-tuning for Recommendation}
Prompt-tuning, initially proposed in NLP~\cite{li2021prefix,chen2022inducer,ding2021openprompt}, has gained traction in recommender systems for its performance improvements, particularly in few-shot scenarios. 
Typically, it involves freezing the backbone model and providing downstream tasks with related tunable prompt embeddings. 
Recently, inspired by NLP works, researchers have introduced prompt-tuning modes to recommender systems, such as encoding user-profiles and behaviors as prompts~\cite{wu2024personalized}, incorporating graph structures as additional prompts~\cite{zhai2023knowledge}, or encoding discrete item features for prompting~\cite{li2023personalized}. 
Further, considering the relevance of few-shot learning and cold-start recommendations,~\cite{wu2023towards,wu2024personalized} have introduced prompt learning to solve the cold-start recommendation problem. 
However, these methods often achieve suboptimal performance in item cold-start scenarios because the corresponding prompt information, such as item profiles, lacks crucial contextual relevance to cold-start recommendations. 
Some research efforts propose leveraging annotated text descriptions and knowledge graphs to activate large language models for recommender tasks~\cite{geng2022recommendation,xu2024prompting,wang2022towards}. 
While effective in igniting the backbone with extra input prompt data, these approaches are less practical in industrial settings due to the high annotation costs associated with millions or billions of new items in such scenarios.

\section{Preliminary}
% This section covers the basic notations and fundamental concepts in \sys. 
% 画图表示不是很清晰，最好用双塔的模式画出来
% 图上的字母物理含义表达不够清晰
% 
\begin{figure*}[t]
\centering
\includegraphics[width=1.0\linewidth]{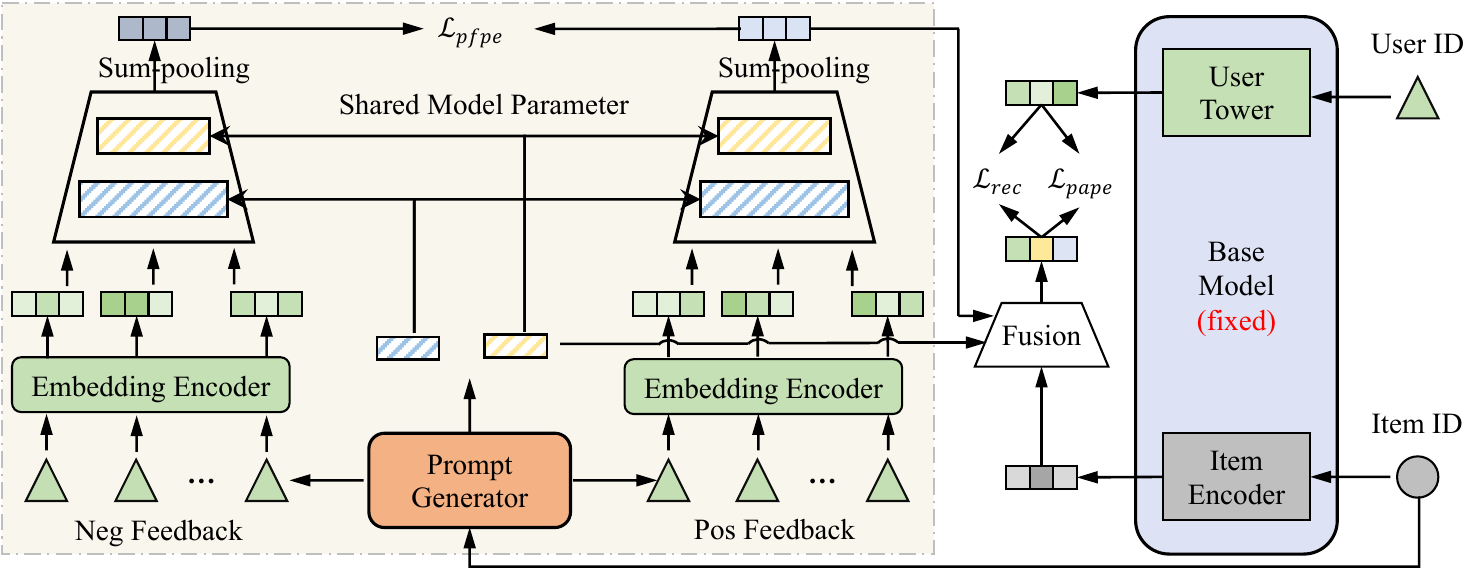}
\caption{Overview of \sys. \sys utilizes item features and pinnacle feedback as the prompt information to generate prompt embedding and personalized prompt network, then \sys optimizes the prompt embedding and eliminates the model bias by the pinnacle feedback prompt-enhance loss and the popularity-aware prompt-enhanced loss separately. }
\label{pipeline}
\end{figure*}

\subsection{Item Cold-start Problem}
%\textbf{\red{Item Cold-start Problem}}. 
We focus on addressing the \emph{item cold-start problem}, where new items have no or few prior interactions, like click or rating logs.
The item cold-start problem requires to provide a scoring function that can accurately estimate the user's preference for a new item, thus provides effectiveness guarantee for the item cold start problem. 

\subsection{Notations \& Problem Formulation}
%\textbf{\red{Problem Formulation}}.
In this work, we use boldface lowercase letters (e.g., $\bm{e}$) to denote vectors, boldface uppercase letters (e.g., $\bm{W}$) to denote
matrices, and calligraphic capital letters (e.g., $\mathcal{D}$) to denote sets. 
Let $\mathcal{U}$,$\mathcal{I}$ denote the sets of users and items, respectively.
The problem here can be formulated as a typical CTR prediction task. 
Each user $u$ $\in$ $\mathcal{U}$ is associated with a set of items $\mathcal{I}^u$ with feedback $y_{ui}$ $=$ $1$ for $i$ $\in$ $\mathcal{I}^u$, indicating that the user has clicked on the items. 
$\mathbf{e}_u, \mathbf{e}_i \in \mathbb{R}^d$ are the input ID embeddings of $u$ and $i$, respectively, $\mathbb{R}$ is the set of real numbers, and $d$ is the embedding dimension. 
In our scenario, given the pair of ($u$, $i$), where $u$ $\in$ $\mathcal{U}$ and $i$ $\in$ $\mathcal{I}$, our goal is to learn a function that can forecast how likely the item will be clicked by the user, i.e. $\hat{y}_{ui}$, $\hat{y}_{ui}$ ranges from 0 to 1, indicating how likely the user will clicked on the item. 

\section{The proposed method}
We initiate the pre-training phase of \sys, showcasing its adaptability across various base models. 
It consists of three parts.

\textit{(1) Embedding Layer.} In our scenario, the input to the base model includes several item content features and categorical ID features, such as item ID. 
These ID features cannot be directly input into the model as they are not trainable for downstream tasks. 
Therefore, we adopt the widely used embedding technique to transform the original sparse features into low-dimensional vectors. 
For user $u$ $\in$ $\mathcal{U}$ and item $i$ $\in$ $\mathcal{I}$, id embeddings $\mathbf{e}_u$ and $\mathbf{e}_i$ are concatenated with their other continuous features to generate the overall embeddings.

\textit{(2) Backbone.} We first introduce the pre-training stage of \sys. Following~\cite{wu2024personalized}, we use the classical SASRec~\cite{kang2018self} as our pre-training model, which can be flexibly substituted with other representation learning methods in recommendations. 
SASRec stacks Transformer~\cite{vaswani2017attention} blocks to encode the historical behavior sequence. 
For each user $u$ with behavior sequence $s_u$, its $l$-layer behavior matrix is formulated as $\mathbf{S}_u^l = (\mathbf{e}_{i_1}^l, \mathbf{e}_{i_2}^l, \ldots, \mathbf{e}_{i_k}^l)$, where $\mathbf{e}_{i_k}^l$ is the $k$-th behavior's representation of $u$ at the $l$-th layer. 
The $(l + 1)$-layer behavior matrix is learned as follows:
\begin{equation} \small
\mathbf{H}_u^{l+1} = \text{Transformer}^l(\mathbf{H}_u^l), \quad \bm{h}_u = f_{\text{seq}}(\mathbf{H}_u^{L},\mathbf{e}_u)
\end{equation}
where $f_{\text{seq}}$ is a MLP encoder, $\bm{h}_u$ is the final user representation of $u$ learned in the pre-training stage, and $L$ is the number of Transformer layers.

Through the base model, we can obtain the hidden representations $\mathbf{h}_u \in \mathbb{R}^d$ and $\mathbf{h}_i \in \mathbb{R}^d$ of user $u$ and item $i$, respectively. 
Then, we can make predictions on the unobserved interaction between user $u$ and item $i$ as follows:
\begin{equation} \small
\hat{y}_{u, i} = \sigma\left(\frac{[\bm{h}_u]^\top \cdot \bm{h}_i}{\tau}\right)
\end{equation}
where $[\cdot]^\top$ represents the transpose of the matrix, $\tau$ represents the temperature coefficient, and $\sigma$ is the sigmoid function.

\textit{(3) Loss.} The objective function used in the base model and \sys is both the negative log-likelihood function defined as:
\begin{equation} \small
\mathcal{L} = -\frac{1}{N}\sum_{(u, i, y) \in \mathcal{D}_{\text{train}}} \left(y_{u, i} \log \hat{y}_{u, i} + (1 - y_{u, i}) \log (1 - \hat{y}_{u, i})\right)
\label{rec}
\end{equation}
where $\mathcal{D}_{\text{train}}$ is the training set, $\hat{y}$ is the output representing the predicted probability whether item $i$ is clicked by $u$, and $y \in {0, 1}$ is the ground-truth label.

\subsection{Prompt Generator}
Existing methods face challenges in effectively prompting recommendation models in cold-start scenarios for two primary reasons. 
Firstly, these methods heavily rely on content features as prompt information. 
These content features may suffer from a semantic gap between their representation and the requirements of recommender tasks, potentially hindering their efficacy. 
Secondly, as experimentally illustrated in Sec.1, we argue that positive feedback information is better suited to serve as prompt information, as it offers contextual information tailored to address the scarcity of positive feedback in item cold-start scenarios. 
From a model perspective, the training samples of popular items, especially the positive samples, provide the majority of training data to the recommender system, resulting in the predominance of training samples from popular items. 
Consequently, this leads to model bias, where the model tends to assign higher scores to popular items, perpetuating their dominance in recommendations.
\subsubsection{Prompt Data}
Initially, we propose utilizing pinnacle feedback as input for prompting, where "pinnacle feedback" refers to users who have provided exceptionally positive feedback for a given item. 
This approach offers advantages on two fronts: (1) These users are crucial for the recommender system to understand the item, and (2) by leveraging user feedback as prompt information and as part of the cold-start item representation, the user-item matching problem can be reframed as a user-user matching problem, potentially increasing the traffic distribution efficiency of cold-start items and aiding more cold-start items in transitioning into popular items. 

To select appropriate positive feedback for item $i$ $\in$ $\mathcal{I}$, we present a criterion to measure the score of each feedback for item $i$, which is also applied in our online recommender system. 
For each user $u$ $\in$ $\mathcal{U}$ who has viewed the item $i$, we consider multiple positive feedback from users (i.e., positive comments and forward) on the items to calculate preference score for selecting pinnacle feedback.
This measurement standard can be flexibly modified to suit different recommender scenarios, e.g., short-video recommendation and e-commerce recommendation. 
Specifically, we measure the sample value $v_{u, i}$ of user (feedback) $u$ to item $i$ using the following formula.
\begin{equation} \small
\label{user_score}
v_{u, i}= \alpha \cdot CR_{u, i} + \beta \cdot IR_{u, i}
\end{equation}
where $CR_{u, i}$ represents the staying time of the user on the item page, and $IR_{u, i}$ is a comprehensive score of user interaction feedback, such as liking the item or following the owner of the item. 
According to the above criteria, we select the top-$k$ users with the highest value as the pinnacle sample list for the item $i$, that is, $Pos_i$ $=$ ($u_1^{pos}$, $u_2^{pos}$, ..., $u_k^{pos}$).

For cases where certain items lack user feedback, referring to ideas behind~\cite{barkan2019cb2cf}, we utilize collaborative filtering information from warm-up items to help cold-start items, we generate pseudo-pinnacle sample information from related popular items. Specifically, for a cold-start item $a$ $\in$ $\mathcal{I}$, we calculate item similarities between itself and popular item $b$ $\in$ $\mathcal{I}$ using the formula:
\begin{equation} \small
v_{b, a}=\frac{\mathbf{h}_{a} \cdot \mathbf{h}_{b}}{\left|\mathbf{h}_{a}\right|^2+\left|\mathbf{h}_{b}\right|^2-\mathbf{h}_{a} \cdot \mathbf{h}_{b}}
\end{equation}
where $\mathbf{h}_{a}$ and $\mathbf{h}_{b}$ represent embeddings of items $a$ and $b$, encoded from the base model. 
We select the most similar popular item to item $a$, and apply Equation~(\ref{user_score}) to extract related positive feedback as pseudo-prompt information.

In order to give full play to the role of the pinnacle samples, we also randomly select $k$ negative feedback of item $i$ (users who do not have any positive interaction with the item $i$), that is, $Neg_i$ $=$ ($u_1^{neg}$, $u_2^{neg}$, ..., $u_k^{neg}$), as the input prompt information. 
Based on the pre-trained \textit{Embedding Layer} in the base model, we use the look-up operation to convert $Pos_i$ and $Neg_i$ into embedding representations $\mathbf{S}_i^{pos}$ $=$ ($\mathbf{e}_{u_1}^{pos}$, $\mathbf{e}_{u_2}^{pos}$, ..., $\mathbf{e}_{u_k}^{pos}$) and $\mathbf{S}_i^{neg}$ $=$ ($\mathbf{e}_{u_1}^{neg}$, $\mathbf{e}_{u_2}^{neg}$, ..., $\mathbf{e}_{u_k}^{neg}$).

\subsubsection{Personalized Prompt Network}
\textit{(1) Model Architecture.} From the model perspective, we use the prompt generator to generate $l$ independent sets of embeddings $\mathbf{e}_i^{p_0}$ $\in$ $\mathbb{R}^{d^{p_0}}$, $\mathbf{e}_i^{p_1}$ $\in$ $\mathbb{R}^{d^{p_1}}$, $\cdots$, $\mathbf{e}_i^{p_l}$ $\in$ $\mathbb{R}^{d^{p_l}}$, for cold start item $i$ $\in$ $\mathcal{I}$, which are different from the \textit{Embedding Layer} in the base model. 
For the specific $n$-th item embedding $\mathbf{e}_i^{p_n}$, we reshape and split them into the weight matrix and bias vector for the personalized prompt network. 
This process for $\mathbf{e}_i^{p_n}$ can be formulated as:
\begin{equation}
\begin{aligned} \small
&\mathbf{e}_i^{p_n} = \mathbf{W}_i^{p_n} // b_i^{p_n}, 
&\left|\mathbf{W}_i^{p_n}\right| + \left|b_i^{p_n}\right| = \left|\mathbf{e}_i^{p_n}\right| = d^{p_n}
\end{aligned}
\end{equation}
where $\mathbf{W}_i^{p_n}$ and $b_i^{p_n}$ denote the weight and bias of $n$-th layer of the personalized prompt networks, $//$ indicates the concatenation operation, $\left| \cdot \right|$ gets the size of the variables.

This process mitigates the problem of model bias, where popular items dominate model updates, as each item only updates its item-personalized prompt networks. 
Furthermore, it ensures that the prompt information fully takes effect, as the personalized prompt network is independent from the original base model and solely encodes the prompt information.

\textit{(2) Complexity.} \sys is parameter-efficient, since it only needs to tune and store a few parameters compared with fine-tuning the base model. 
The tunable parameters in \sys only cost in the personalized prompt network part ($\mathbf{W}_i^{p_n}$ and $b_i^{p_n}$), which is very efficient. 
In \sys, the number of parameters to be updated merely 20.1\%, 27.6\%, 17.7\% and 25.6\% of those fine-tuning in MovieLens100K, MovieLens1M, KuaiRand, and TMall datasets. 

\subsection{Prompt Learning in \sys}
In the previous section, we obtain the positive feedback list $\mathbf{S}_i^{pos}$, negative feedback list $\mathbf{S}_i^{neg}$ and the parameters of the personalized prompt network for the cold-start item $i$ $\in$ $\mathcal{I}$ based on the prompt generator. 
Next, $\mathbf{S}_i^{pos}$ and $\mathbf{S}_i^{neg}$ are fed into the personalized prompt network and the feature interaction is realized via the concatenation of the output of each layer. Take $\mathbf{S}_i^{pos}$ as an example: 
\begin{equation}
\begin{aligned}
\small
\mathbf{h}_l^{pos} &= \sigma \left(\mathbf{W}_i^{p_{l - 1}} \otimes (\cdots (\mathbf{W}_i^{p_{0}} \otimes \mathbf{h}_{0} + b_i^{p_{0}}) + b_i^{p_{l - 1}}\right),
\end{aligned}
\end{equation}
where $\otimes$ denotes the matrix multiplication, $\sigma$ indicates the activation function, $l$ represents the layers of the personalized prompt network, $\mathbf{h}_l^{pos}$ is the output of the prompt data fed through the prompt network, and $\mathbf{h}_0^{pos} = \mathbf{S}_i^{pos}$. 
We can also get $\mathbf{h}_l^{neg}$ for $\mathbf{S}_i^{neg}$.

To fully leverage the potential of prompt information in \sys, we introduce the pinnacle feedback prompt-enhanced loss $\mathcal{L}_{pfpe}$ and the popularity-aware prompt-enhanced loss $\mathcal{L}_{pape}$.

\textbf{Per-sample Pinnacle Feedback Prompt-enhanced Loss.}
Per-sample Positive Feedback Prompt-enhanced Loss is designed to fully take advantange of high-value positive samples for cold-start items. 
To achieve this, for the item $i$ $\in$ $\mathcal{I}$, we leverage the concept of contrastive learning to widen the gap between its pinnacle and negative feedback representations. 
We calculate the sum of the distances between each pinnacle feedback and each negative feedback of $i$ as follows:
\begin{equation} \small
\Delta_i = \sum_{x \in Pos_i} \sum_{y \in Neg_i} \left\|\mathbf{h}_{l, x}^{pos} - \mathbf{h}_{l, y}^{neg}\right\|
\end{equation}
where we employ the $L_1$ distance to calculate the discrepancy between each pinnacle feedback vector and each corresponding negative feedback vector. 
Furthermore, we utilize the log function to implement the loss:
\begin{equation} \small
\mathcal{L}_{pfpe} = \log \left(1 + exp\left(- \Delta_i\right)\right).
\end{equation}
Optimizing $\mathcal{L}_{pfre}$ enables \sys to indirectly memorize valuable feedback information in both the prompt network parameters and the prompt embedding. 
Additionally, the pair-wise $\mathcal{L}_{pfre}$ supplements order information to the base model, which is trained in a point-wise manner and lacks ordering ability.

\textbf{Intra-batch Popularity-aware Prompt-enhanced Loss.} 
The final representation of the item $i$ $\in$ $\mathcal{I}$ is composed of the output $\mathbf{h}_i$ of the base model, the average of its positive feedback list $\mathbf{S}_i^{pos}$, and the last layer of personalized prompt network embedding $\mathbf{e}_i^{p_l}$: 
\begin{equation} \small
\mathbf{e}_i^{final} = MLP\left(\mathbf{h}_i, \frac{1}{k}\sum\mathbf{S}_i^{pos}, \mathbf{e}_i^{p_l}\right)
\end{equation}
where MLP is the Multi-layer Perceptron network. 
With the final embeddings, we adopt a simple but widely-used inner product model, i.e., $\hat{y}_{u, i}$ $=$ $\left[\mathbf{e}_u^{final}\right]^\top \cdot \mathbf{e}_i^{final}$, to estimate the value of $\hat{y}_{u, i}$, which is the interaction probability between a given pair of user and item. 

Industrial recommender systems frequently exhibit biased preferences for popular items. 
This bias emerges because popular items contribute the majority of positive samples.
Consequently, as depicted in Figure~\ref{case2} (b), the estimated scores for cold-start items in positive samples are even lower than those for popular items in negative samples. 
To address this, \sys introduces an intra-batch popularity-aware prompt-enhanced loss. Before that, for the samples (pairs of the user $u$ and the item $i$) in a batch, we can get two sets $\mathcal{D}{{pos\_cold}}$ and $\mathcal{D}{neg\_warm}$ respectively based on the cold-start status of the item and the label. 
Here, $\mathcal{D}{pos\_cold}$ is the positive sample set of cold-start items within a batch, and $\mathcal{D}{neg\_warm}$ is the negative sample set of popular items within the batch. 
We calculate the distance between positive cold-start samples and negative popular samples within the batch and use the following loss to further push the distance to achieve fair scoring.
\begin{equation} \small
\Delta_{batch} = \sum_{(\bar{u}, \bar{i}) \in \mathcal{D}_{pos\_cold}} \sum_{(\widetilde{u}, \widetilde{i}) \in \mathcal{D}_{neg\_warm}} 
 \left(\hat{y}_{\bar{u}, \bar{i}} - \hat{y}_{\widetilde{u}, \widetilde{i}}\right)
\end{equation}
\begin{equation}
\mathcal{L}_{pape} = \log \left(1 + exp\left(- \Delta_{batch}\right)\right)
\end{equation}

\subsection{Optimization}
% To k, We optimize the Equation~(\ref{rec}) as the main objective.
% The difference is that $\hat{y}$ is calculated with the final representations. 
To fully leverage knowledge from the pretrained model and \sys, we directly fuse the representations from the pretrained model and \sys by concatenating the embeddings together, thus obtaining the final embeddings. 
The final prediction value is estimated from the final embedding. 
The overall objective function is formulated as:
\begin{equation}\small
\mathcal{L} = \lambda_1\mathcal{L}_{pfre} + \lambda_2\mathcal{L}_{pape} + \mathcal{L}_{rec}
\end{equation}
where $\lambda_1$ and $\lambda_2$ are positive coefficients serving as balancing factors for the multiple loss functions.
$\mathcal{L}{rec}$ has the same format as Equation \ref{rec}, which is important to inject representations with personalized recommender knowledge. 
Following the training paradigm in prompt learning~\cite{lester2021power}, the base model is frozen during the optimization. 
Only tuning on the small set of model parameters without changing the base model is not only parameter-efficient, but it can also better leverage the pre-trained model and avoid the catastrophic forgetting problem~\cite{kirkpatrick2017overcoming}.

\section{Experiment}
In this section, we conduct an empirical assessment of our framework and present comprehensive results. Our focus lies on addressing the following six questions:
\textbf{Q1:} Can \sys surpass state-of-the-art baseline methods in terms of recommendation performance in cold-start scenarios across real-world datasets?
\textbf{Q2:} What are the advantages of the prompt learning module within \sys for enhancing downstream recommender tasks?
\textbf{Q3:} How effective is the proposed positive feedback prompt information? Can it be replaced by other types of prompt information?
\textbf{Q4:} How does the personalized prompt network affect the overall system performance?
\textbf{Q5:} Does the prompt paradigm in \sys effectively enhance the high-value positive feedback information?
\textbf{Q6:} Does \sys effective in providing high-quality recommender results in the industrial application? 

\subsection{Datasets \& Baselines}
\paragraph{Datasets}
We utilize four public recommender datasets - MovieLens 100K \& 1M~\cite{harper2015movielens} about movie recommendations, KuaiRand~\cite{gao2022kuairand} of short-video recommendations and TMall~\cite{TMall} of E-commerce recommendations, to evaluate the effectiveness of our methods. 
Specifically, we preprocessing the datasets for the evaluation on cold-start performance. 
% And the detailed description about these preprocessing process of datasets is in Appendix A.1. 
For all four datasets, We divide items into two groups, popular and cold-start based on their positive feedback interaction frequencies, where items with more than positive feedback interactions are popular and others are cold-start. We use of 20, 50, 50 and 20 for MovieLens 100K, MovieLens 1M, KuaiRnad-Pure and Tmall dataset, respectively. Note
that the ratio of cold-start items to  popular items is approximately 8:2, which
is similar to the definition of long-tail items.
Following the same leave-one-out technique in existing works~\cite{wu2024personalized,zhang2023empowering}, we  take the last interactive item for each user as the test data, the second-to-last as validation data, and the remaining data as training data to simulate the online recommendation environment. 
The detailed statistics are summarized in Table~\ref{Dataset}.
\begin{table}[t]
    \caption{Overview of the dataset.}
    \label{Dataset}
    \resizebox{\columnwidth}{!}{%
    \begin{tabular}{cccccc}
    \toprule
    \multirow{2}{*}{\textbf{Dataset}} & \multirow{2}{*}{\textbf{\#Users}} & \multirow{2}{*}{\textbf{\#Items}} & \multirow{2}{*}{\textbf{\#Ratings}} & \multicolumn{2}{c}{\textbf{\#Features}} \\ 
   \cmidrule(lr){5-6}  & & & & \textbf{User} & \textbf{Item} \\ \midrule
    MovieLens 100K & 943   & 1,682 & 100,000   &  23 & 18 \\
    MovieLens 1M   & 6,040  & 3,706 & 1,000,209  & 23 & 18 \\
    KuaiRand-Pure  & 27,285 & 7,583 & 2,599,187 & 25 & 55\\ 
    TMall      & 52,797  & 22,955 & 6,330,878    &  2 & 3   \\\bottomrule
    \end{tabular}%
    }
\end{table}

\paragraph{Baselines.}
In order to comprehensively assess the efficacy of the prompt tuning mode in our system (\sys), we conduct a comparison with two approaches: Pre-train, where we directly apply the pre-trained model (SASRec~\cite{kang2018self}) to the test set, and Fine-tuning, which involves tuning all parameters of the pre-trained model during the tuning phase.
Besides, we compare \sys with four popular and effective CF model: DCN~\cite{wang2017deep}, DeepFM~\cite{guo2017deepfm}, SASRec~\cite{kang2018self}, DSSM~\cite{huang2013learning}, four state-of-the-art cold-start recommender models: CDN~\cite{zhang2023empowering}, DropoutNet~\cite{volkovs2017dropoutnet}, CB2CF~\cite{barkan2019cb2cf}, MetaEmb~\cite{pan2019warm}, and two prompt-tuning in recommender systems: PPR~\cite{wu2024personalized}, PLATE~\cite{wang2023plate}. 
Our code for \sys is available at \url{https://github.com/PROMOREC/PROMO}. 
% More details of our baseline methods can be found in Appendix A.2.
The main characteristic of all baselines are listed below:

\begin{table*}[t]

    \caption{Overall Performance Comparison in Item Cold-Start Scenario. The best-performing results are highlighted in bold, while the second-best performance is indicated with underlines.}

    \label{end2end}
    \resizebox{\textwidth}{!}{%
    \begin{tabular}{c cccc cccc cccc cccc}
    \toprule
    \multirow{2}{*}{\textbf{Models}} & \multicolumn{4}{c}{\textbf{MovieLens 100K}}  & \multicolumn{4}{c}{\textbf{MovieLens 1M}} & \multicolumn{4}{c}{\textbf{KuaiRand}} & \multicolumn{4}{c}{\textbf{TMall}}   \\ \cmidrule(lr){2-5} \cmidrule(lr){6-9} \cmidrule(lr){10-13}\cmidrule(lr){14-17} 
             & \textbf{H@5}  & \textbf{H@10}   & \textbf{N@5} & \textbf{N@10}   & \textbf{H@5}  & \textbf{H@10}  & \textbf{N@5} & \textbf{N@10}   & \textbf{H@5}  & \textbf{H@10}   & \textbf{N@5} & \textbf{N@10}   & \textbf{H@5}  & \textbf{H@10}  & \textbf{N@5} & \textbf{N@10}  \\ \midrule
    Pre-train       & 14.6            & 26.2                    &  9.3            & 13.3                     & 27.7            & 42.5                     & 18.1 
    & 22.2                       & 13.3            & 25.8                       & 7.9           & 11.9                    & 3.6            & 7.9                    & 2.2            & 3.6          \\
    Fine-tune       & 17.4            & 28.2                    &  10.7            & 14.2                     & 28.1            & 42.7                     & 18.3            & 22.8                       & 24.9            & 47.5                       & 16.6            & 32.5                    &6.1             & 16.7                    &9.8             & 8.3          \\
    CDN       & 17.6            & 27.4                    &  11.6            & 14.7                     & 17.5            & 30.4                     & 10.8            & 14.9                       & 22.2            & 35.5                       & 14.2            & 18.5                    & 2.6            & 4.9                    & 1.7            & 2.4          \\
    DCN       & 15.0            & 29.3                    &  9.1            & 15.4                     & 25.6            & 40.9                     & 16.4            & 21.3                       & 20.4            & 32.3                       & 13.3            & 17.2                    & 7.8            & 13.7                    & 5.0            & 6.8          \\
    DeepFM             & \underline{17.6}           & 28.0                       & \underline{11.3}            & 14.6                        & 30.5            & 45.7                      & 20.2            & 25.1                        & 24.1            & 35.5                        & 16.7            & 20.4                      & 4.8            & 10.6                       & 2.7            & 4.5         \\
    SASRec                & 17.4            & 30.2                         & 10.8            & 14.9                       & \underline{56.9}            & \underline{69.2}                        & \underline{43.4}            & \underline{47.6}                       & 82.5            & 89.3                     & \underline{70.5}            & 72.7                      & \underline{23.1}            & \underline{33.5}                     & \underline{15.8}            & \underline{19.6}          \\
    % DSSM               & \underline{19.2}            & 30.7                    & \underline{12.8}            & 16.5                        & 55.3            & \underline{69.6}                      & 41.1            & 45.7                    & \underline{83.7}            & \underline{90.1}                       & \underline{71.1}            & 73.2                      & 21.9            & 32.3                      & 15.0           & 18.5          
    DSSM               & 16.4            & 29.1                    & 10.7           & 15.1                        & 28.8           & 34.7                      & 18.8            & 23.6                   & 22.3            & 34.2                       & 14.8            & 18.6                     & 21.9            & 32.3                      & 15.0           & 18.5 \\
    DropoutNet             & 16.1            & 28.6                        & 10.2            & 14.2                      & 32.2            & 46.9                       &  21.1           & 25.8                      & 24.9            & 40.7                      & 15.4            & 20.5                       & 6.4            & 11.2                      & 4.6            & 6.2          \\
    CB2CF                & 15.7            & 26.4                       & 9.9            & 13.1                  & 19.7            & 34.0                     & 11.9            & 16.8                      & 17.3            & 31.3                    & 10.3            & 14.8                 & 3.2           & 6.1                   & 1.9            & 2.8          \\
    MetaEmb             & 17.1            & 29.3                        & 10.9            & 14.9                       & 30.3            & 44.8                        & 20.2           & 24.8                     & 16.6            & 30.0                     & 10.0            & 14.3                    & 9.0            & 17.6                       & 5.3            & 8.0          \\
    PPR           & 10.5            & \underline{32.8}                         & 6.1            & \underline{17.6}                      & 55.9            & 69.0                 & 41.9            & 46.1                         & \underline{83.1}            & \underline{90.0}                        & 70.3            & \underline{72.7}                    & 22.5            & 32.3                      & 15.5            & 18.7          \\
    PLATE              & 15.7            & 26.0                      & 9.9            & 13.2                  & 24.2            & 39.1                    & 15.3            & 20.1                   & 17.5            & 30.3                      & 10.8            & 15.0                      & 3.7           & 7.5                     & 2.2            & 3.4        \\ \midrule
\sys                & \textbf{33.7}            & \textbf{43.3}                       & \textbf{25.5}            & \textbf{28.6}                       & \textbf{57.5}            & \textbf{70.6}                      & \textbf{43.8}            & \textbf{48.0}                         & \textbf{88.7}            & \textbf{92.6}                      & \textbf{80.3}            & \textbf{81.6}                    & \textbf{24.2}            & \textbf{34.2}                       & \textbf{17.0}            & \textbf{20.2}          \\ \bottomrule
    \end{tabular}%
    }
    \end{table*}

\label{baseline_methods}
\begin{itemize}[leftmargin=*]
    \item CDN~\cite{zhang2023empowering}: CDN presents a Cross Decoupling Network (CDN) to enhance long-tail item recommendation by addressing biases in user preference predictions.
    \item DCN~\cite{wang2017deep}: DCN is a novel cross network that is efficient in learning certain bounded-degree feature interactions. As it is poweful and the feature crossing manner is similar to prompt network in \sys, we take it as one of the baseline method. 
    \item DeepFM~\cite{guo2017deepfm}: DeepFM derives an end-to-end learning model that emphasizes both low- and highorder feature interactions by a shared input to the “wide” and “deep” parts.
    \item SASRec~\cite{kang2018self}: SASRec capture both long-short recommender interests by utilizing both long-term semantics (like an RNN) and using an attention mechanism to makes its predictions based on relatively few actions. 
    \item DSSM~\cite{huang2013learning}: SASRec capture both long-short recommender interests by utilizing both long-term semantics (like an RNN) and using an attention mechanism to makes its predictions based on relatively few actions. 
    \item DropoutNet~\cite{volkovs2017dropoutnet}: DropoutNet modify the learning procedure to explicitly condition the model for the missing input, and train DNNs to generalize to missing input.
    \item CB2CF~\cite{barkan2019cb2cf}: CB2CF introduces the model for bridging the gap between items’ CB profiles and their CF representations. It is supervised by CF information, produces significantly better results than classical CB models that use the same CB data.
    \item MetaEmb~\cite{pan2019warm}: MetaEmb trains generators by making use of previously learned ads through gradient-based meta-learning.
    \item PPR~\cite{wu2024personalized}: PPR introduces prompt to pre-trained recommendation models for cold-start recommendation.
    \item PLATE~\cite{wang2023plate}: PLATE conducts prompt tuning with two novel prompt modules, capturing the distinctions among various domains and users. 
\end{itemize}

\subsection{Evaluation Metrics \& Parameter Settings}

For each user in the test set, we take all the items that the user has not interacted with as negative samples. 
For each testing user-item positive pair, we randomly sample 100 items which are not interacted by the user to generate negative pairs for evaluation, which mirrors similar settings in existing works.
We evaluate the proposed \sys with two metrics: Hitrate@K and NDCG@K. 
These metrics can be formulated as follows:
\begin{equation} \small
\label{equ:hr}
Hitrate@K = \frac{ \sum_{u} \sum_{j=1}^{K}rel_j }{\sum_{u}|T(u)|}
\end{equation}
\begin{equation} \small
NDCG@K = \frac{1}{|u|} \sum_{u} \frac{DCG@K}{IDCG@K}, DCG@K=\sum_{j=1}^{K}\frac{2^{rel_j}-1}{log_2(j+1)}
\end{equation}
where $R(u)$ and $T(u)$ represent the model recommendation set and the whole test set respectively; $rel_j=0/1$ indicates item at rank $j$ in $R(u)$ is also belonging to $T(u)$. IDCG is the DCG score for the most ideal ranking, which is ranking the items top down according to their real score. 
In our experiment, $K$ is separately set to be 5 and 10. A higher score indicates a model with better retrieval or ranking ability.

\subsection{End-to-End Comparison} 
To address \textbf{Q1}, we present the performance results of our end-to-end comparison in Table~\ref{end2end}. 
Across four recommender datasets, our system (\sys) consistently outperforms all baseline methods across various metrics (H@1, H@5, H@10, N@1, N@5, and N@10). 

Compared to cold-start baseline methods, \sys achieves superior performance. 
For instance, \sys outperforms CB2CF and MetaEmb by large margins of 37.8\% and 27.2\%, respectively, in HitRate@5 on the MovieLens 1M dataset. 
This highlights the importance of positive feedback in enhancing recommender tasks compared to content-based information. 
Moreover, CF-based methods generally achieve better performance than content-based methods, further underscoring the effectiveness of utilizing feedback information for recommendations. 
Despite this, \sys achieves significant improvements over CF-based methods, e.g., surpassing SASRec by 16.3\%, 13.1\%, 14.7\%, and 13.7\% according to the metrics of H@5, H@10, N@5, and N@10, respectively, on the MovieLens 100K dataset. 
This superiority indicates the effectiveness of the proposed pinnacle feedback information for recommendations.

Additionally, it is noteworthy that the cutting-edge prompt recommender method PPR often achieves the second-best performance. 
This underscores the effectiveness of the prompt paradigm in addressing few-shot learning problems. 
However, \sys surpasses cutting-edge prompt recommender methods like PPR and PLATE across all metrics. 
While PPR adopts SASRec as its backbone model and follows a two-stage training approach, \sys's superiority is evident due to several factors. 
PPR's reliance on content features for prompt information may lead to suboptimal results due to the lack of utilization of positive feedback information. 
Furthermore, PPR's use of a shared neural network to encode prompt information may introduce model bias, favoring popular items over cold-start ones.

\subsection{Interpretability of \sys}
\begin{figure}
\centering
    \subfigure[Item representations from DSSM.]{
    \centering
    \includegraphics[width=0.47\linewidth]{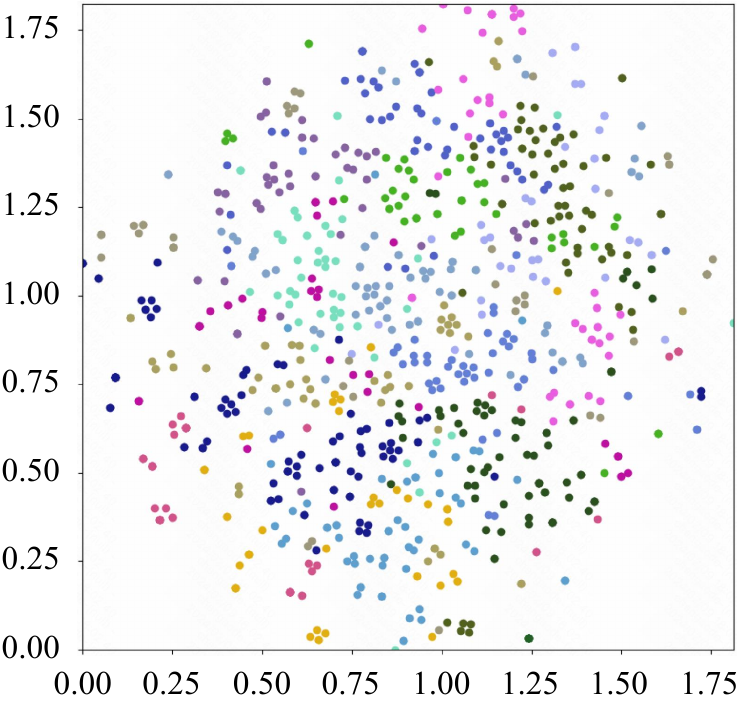}
    }
    \subfigure[The item representations obtained after training in the proposed \sys.]{
    \centering
    \includegraphics[width=0.47\linewidth]{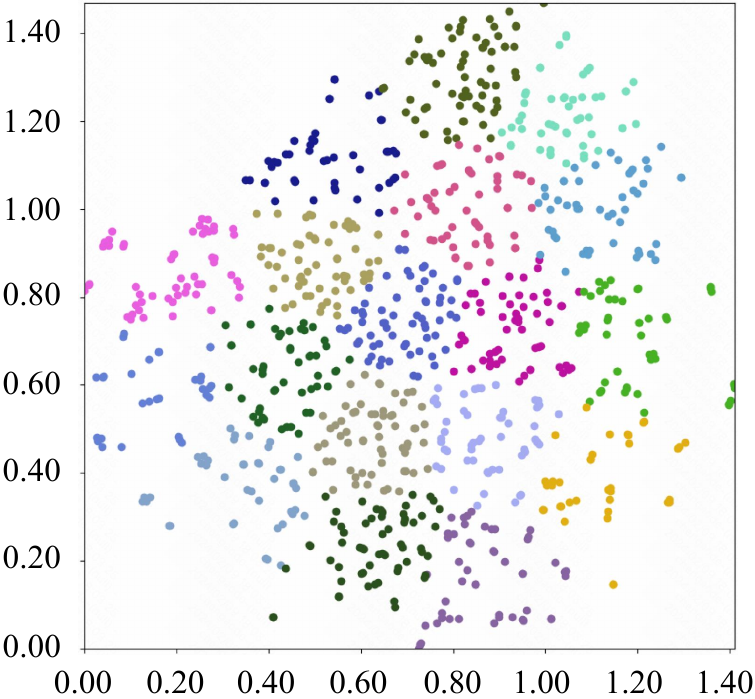}
    }
    \caption{The visualization of item representation involves the assignment of class labels to nodes through the application of K-means clustering on the original input data. 
    The nodes are differentiated into various classes through the use of distinct colors. 
    The representation of nodes is visualized utilizing t-SNE~\cite{van2008visualizing}.}

    \label{visualization}
\end{figure}

To elucidate why \sys is effective for downstream recommender tasks to answer Q2, we begin by revaluating the architecture of \sys. 
As delineated in the model overview, \sys operates as a flexible module within a two-stage framework, with the base model held constant. 
We contend that \sys is more efficacious due to the prompt embedding it generates containing aligned information pertinent to the downstream tasks. 
This alignment stems from the informative nature of pinnacle positive feedback, which guides the recommender system in item distribution. 
To illustrate this phenomenon, we visualize the item representations produced by \sys and a baseline of DSSM model. 
Specifically, upon convergence on the MovieLens100K dataset, we extract the final output embeddings of items from both models and visualize all item embeddings. 
The left segment of Figure~\ref{visualization} depicts the node representations of the DSSM, while the right segment illustrates the embeddings from \sys. 
It becomes apparent that items sharing the same classes are more closely mapped together in \sys, whereas the base model fails to implicate category information. 
This observation suggests that recommender systems, when better aligned with the recommendation tasks, can enrich representations with collaborative relations concerning content information, such as item class information in this instance. 
These learned content relations can furnish pertinent information aiding in task optimization in an end-to-end fashion. 
Consequently, the prompt learning module enhances the model's capacity to produce discriminative item representations by assimilating more task-relevant information from the proposed prompt optimization method in \sys, thereby augmenting the performance of downstream tasks.

\subsection{Influence on Prompt Information}
\label{sec:prompt-info}
\begin{table}[]
\caption{The test accuracy of variants of \sys to examine the effectiveness of pinnacle feedback prompt information.}
\label{tab:promptInfo}
\resizebox{\columnwidth}{!}{%
\begin{tabular}{c cccc cccc cccc}
\toprule
\multirow{2}{*}{}      & \multicolumn{2}{c}{\textbf{MovieLens 100K}}                    & \multicolumn{2}{c}{\textbf{MovieLens 1M}}     & \multicolumn{2}{c}{\textbf{KuaiRand}}            \\ \cmidrule(lr){2-3} \cmidrule(lr){4-5}  \cmidrule(lr){6-7}
                       & \textbf{H@10} & \textbf{N@10}  & \textbf{H@10} & \textbf{N@10} & \textbf{H@10} & \textbf{N@10} \\ \midrule
\sys-I   & 34.8   & 18.3 & 68.8     & 45.9 & 90.0 & 72.9 \\
\sys-F & 35.1   & 17.6 & 69.1      & 46.1  & 90.1 & 73.1\\
\sys-IF   & 35.7   & 18.7 & 69.2     & 46.3 & 90.2 & 73.0 \\ \midrule
\sys       & \textbf{43.3}   & \textbf{28.6}       & \textbf{70.6}      & \textbf{48.0}  & \textbf{92.6} & \textbf{81.6}      \\ \bottomrule
\end{tabular}%
}

\end{table}
To address \textbf{Q3}, we conducted experiments to evaluate the effectiveness of the proposed prompt information. 
\sys represents the first work to utilize pinnacle feedback as prompt information. 
To assess the efficacy of this approach, we replaced pinnacle feedback with commonly used prompt information from other prompt recommendation approaches. 
Specifically, we replaced the pinnacle feedback prompt information with item-IDs (referred to as \sys-I), item features (referred to as \sys-F), and both IDs and item features (referred to as \sys-IF), while keeping other components of \sys unchanged. 
We evaluated the effectiveness of prompt information on the MovieLens 100K, MovieLens 1M, and KuaiRand datasets, with results presented in Table~\ref{tab:promptInfo}.

It is evident that \sys outperforms all its variants according to the HitRate and NDCG metrics. 
This indicates that the prompt information proposed in \sys is more relevant in indicating user interests, thereby enhancing downstream tasks. 
Additionally, it's noteworthy that \sys-I exhibits the poorest performance among the variants. 
This can be attributed to the fact that cold-start items scarcely provide samples for the recommender system, resulting in their ID embeddings being inadequately updated, thereby failing to provide sufficient information for model prediction. 
Although \sys-IF achieves considerable performance compared to other \sys variants, it still falls behind \sys. 
For example, it lags behind by 10.6\% and 11.9\% according to H@10 and N@10 on MovieLens 100K, and falls short by 2.4\% and 8.6\% on KuaiRand according to H@10 and N@10 respectively. 
This underscores the significance of positive feedback information in recommendations, especially in cold-start scenarios.

\subsection{Influence of Personalized Prompt Network}
\begin{table}[]
\caption{The performance on MovieLens 100K, MovieLens 1M and KuaiRand to illustrate the effectiveness of personalized prompt network in \sys.}
\label{tab:promptNet}
\resizebox{\columnwidth}{!}{%
\begin{tabular}{c cccc cccc}
\toprule
\multirow{2}{*}{}      & \multicolumn{2}{c}{\textbf{MovieLens 100K}}                    & \multicolumn{2}{c}{\textbf{MovieLens 1M}}     & \multicolumn{2}{c}{\textbf{KuaiRand}}            \\ \cmidrule(lr){2-3} \cmidrule(lr){4-5}  \cmidrule(lr){6-7}
                       & \textbf{H@10} & \textbf{N@10}  & \textbf{H@10} & \textbf{N@10} & \textbf{H@10} & \textbf{N@10} \\ \midrule
\sys-M & 35.7   & 22.9 & 69.4      & 47.6 & 90.3 & 76.6  \\
\sys-T  & 28.7   & 14.6 & 62.8      & 42.9  & 89.4 & 73.1 \\ \midrule
\sys   & \textbf{43.3}   & \textbf{28.6}       & \textbf{70.6}      & \textbf{48.0}  & \textbf{92.6} & \textbf{81.6}      \\ \bottomrule
\end{tabular}%
}

\end{table}
To address \textbf{Q4}, we assessed the effectiveness of the personalized prompt network by comparing the performance of related variants of \sys. 
Specifically, \sys was evaluated in the following configurations: (i) substituting the personalized prompt network with a global shared MLP prompt network (referred to as "\sys-M"), and (ii) removing the personalized prompt network module and incorporating the positive feedback information directly as an input feature to the base model, denoted as "\sys-T". The performance results are provided in Table~\ref{tab:promptNet}.

It is evident that \sys-M outperforms \sys-T across all metrics in the three datasets. 
This superiority is attributed to the exclusive prompt network in \sys-M, which emphasizes the importance of pinnacle feedback, thereby avoiding any reduction in effectiveness that may occur when combined with other features in \sys-T.
However, \sys-M encounters the model bias problem, as the model parameters become dominated by popular items, leading to suboptimal performance in cold-start recommendations. 
On the contrary, \sys overcomes this issue with the personalized prompt network, which enhances the pivotal pinnacle feedback information while simultaneously avoiding the model bias problem.

\subsection{Effectiveness of Pinnacle Feedback Enhancement} 
\begin{figure}[t]
\centering
\includegraphics[width=\linewidth]{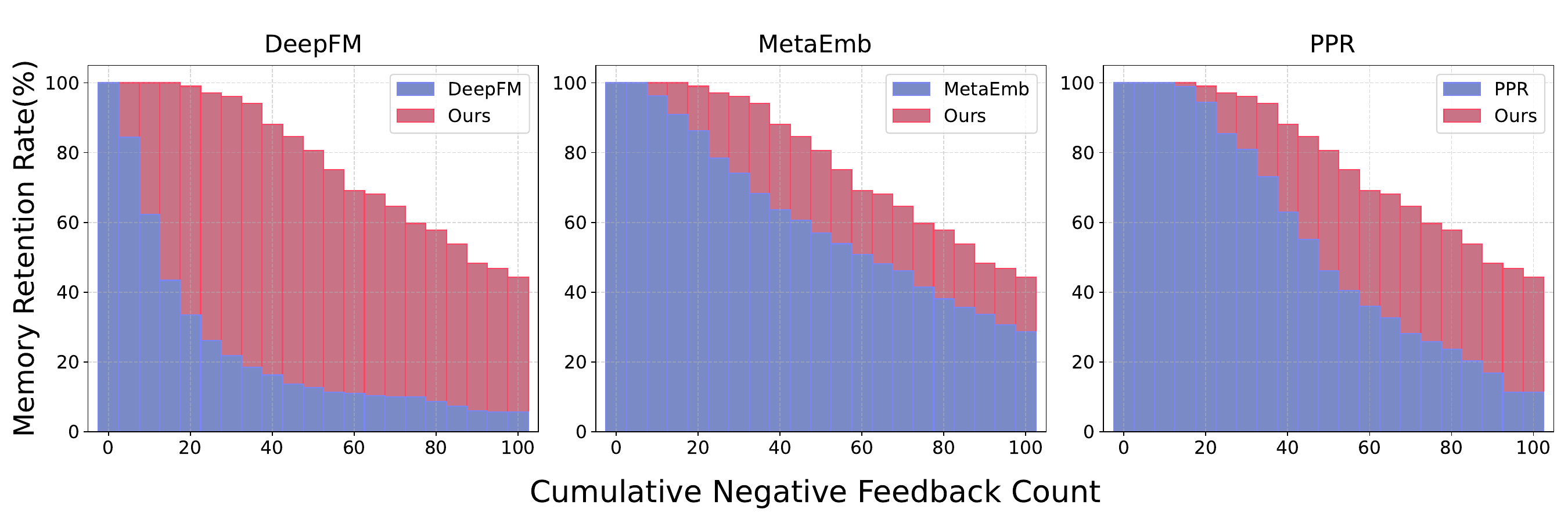}
\caption{The memory retention rate corresponds with the accumulative negative feedback amount on MovieLens 1M dataset. Each bar represents the memory retention rate correponds to a certain negative feedback amount. }
\label{observation_neg}

\end{figure}

To address \textbf{Q5}, we evaluated whether the pinnacle feedback information is effectively retained in \sys. 
We defined a memorized pinnacle feedback as a pinnacle sample accurately estimated by the model at time $t_0$ as well as at $t_1$. 
The memory retention rate was calculated as the memorized pinnacle feedback amount at $t_1$ compared to the total pinnacle amount truly estimated at $t_0$.

We randomly selected 500 pinnacle feedback samples and assessed the memory retention rate with increasing accumulation of negative feedback on the corresponding items. 
For comparison, we selected three representative baseline methods due to page limitations: DeepFM from CF-based methods, MetaEmb from content-based methods, and PPR from prompt learning methods. The memory retention rate comparison is depicted in Fig.~\ref{observation_neg}.
It was observed that as the cumulative negative feedback amount increased, the memory retention rate declined in all three baseline methods. 
The decline was more pronounced with higher cumulative negative feedback amounts, particularly evident in the DeepFM baseline. 
This phenomenon can be attributed to CF-based methods' focus on recent feedback information, leading them to gradually forget earlier samples under online streaming training modes, resulting in the forgetting of original pinnacle feedback information.
In comparison, \sys achieved a high memory retention rate, demonstrating its effectiveness in memorizing pinnacle feedback information and facilitating recommender performance.

\subsection{Production A/B Test} 
To answer \textbf{Q6}, we assess the performance of \sys through its deployment on a popular short-video sharing platform in China - where users can upload their short-videos and enjoy short-videos by other users, using the standard A/B testing methodology.

%, we conducted a comparison to evaluate its efficacy.
\begin{table}[t]
    \centering
    \caption{Online cold-start recommendation evaluation.}

    \resizebox{\columnwidth}{!}{%
    \begin{tabular}{ccccc}
        \toprule
         & Click Rate & Video Play Time & Video Like&  Video Collecting \\\midrule
        \sys & +3.2 \% & +4.8\% & +3.9\% & +4.0 \% \\
        \bottomrule
    \end{tabular}%
    }
    
    \label{coldResult}

\end{table}

We randomly divided users into two groups for online evaluation, with each group comprising more than 30 million users to ensure statistical significance. 
The sole distinction between these groups lies in the cascading process for online serving: users in the first group experience the recommender system that incorporates \sys, while users in the second group are exposed to a online baseline method, which is similar to the SASRec baseline method.

The online serving performance is evaluated using metrics that consider user engagement, such as Click Rate, Video Play Time, Video Like and Video Collecting.
The first two metrics reflect the user's implicit satisfaction, while the latter two reflect the user's explicit preference expressed through behavior. 

The comparison was monitored over 14 consecutive days, and the average performance for item cold-start recommendations is presented in Table~\ref{coldResult}.
We observed a consistent growth trend in both the implicit feedback and explicit feedback.
For example, \sys gain the explicit user feedback on video like of 3.9\% lifting, along with the implicit user feedback of videp play time of 4.8\% lifting. 
demonstrates that \sys provides users with more satifactory recommender results. 
The online evaluation in cold-start recommendation settings highlights the effectiveness of our method.
% and versatility of our proposed method.

\section{Conclusion}
Item cold-start challenge plays a pivotal role in the success of online recommender systems. 
% Prompt learning, a robust technique from NLP used to address zero- or few-shot problems, has been adapted for recommender systems to tackle similar hurdles. 
Prompt learning, has been adapted for recommender systems to tackle similar hurdles. 
% However, prevailing methods often rely on content-based properties or text descriptions for prompting, potentially falling short due to semantic gaps with downstream tasks. 
However, prevailing methods often rely on content-based properties or text descriptions for prompting, potentially falling short due to semantic gaps with downstream tasks. 
We advocate for leveraging positive feedback as a more task-relevant prompt for recommendations.
% given its significance in both content-based and collaborative-based recommendations. 
This is especially crucial for cold-start items, which grapple with limited positive feedback. 
Moreover, the dominance of positive feedback from popular items introduces model bias, where these items receive higher scores from the recommender system.
To combat these challenges, we propose leveraging high-value positive feedback, referred to as pinnacle feedback, as prompt information. 
Our investigation into the efficacy of pinnacle feedback as prompt data for cold-start items, along with the development of a prompt generator to acquire both pinnacle and negative feedback prompt information. 
Furthermore, we introduce item-wise personalized prompt networks to mitigate model bias. 
Extensive experiments on four real-world datasets demonstrate the superiority of our model over state-of-the-art methods. 
Additionally, \sys has been successfully deployed on a industrial recommender system in China, a billion-user scale commercial short-video application, achieving remarkable performance gains across various commercial metrics within cold-start scenarios.

\begin{acks}
% We thank the anonymous reviewers for their valuable suggestions. 
This work is partially sponsored by NSFC (62032003).
\end{acks}

\bibliographystyle{ACM-Reference-Format}
\bibliography{sample-base}

% \appendix

\end{document}